\documentclass[12 pt,twoside,aps,prd,amsmath,amssymb,
tightenlines,showpacs,showkeys,eqsecnum]{revtex4}
\usepackage{mathptmx}
\usepackage{bm}
\usepackage{graphicx}
\usepackage{hyperref}
\newcommand {\ve}{\varepsilon}

\newcommand {\bp}{\bar \psi}

\newcommand {\p}{\psi}

\def \myfigures #1#2#3#4#5#6#7#8
{\begin{figure}[ht]
    \begin{center}
        \includegraphics[width=#2 \textwidth, angle=-90]{#1.eps}
        \hfill
        \includegraphics[width=#6 \textwidth, angle=-90]{#5.eps}
        \parbox[t]{#4\textwidth}{\caption {#3}\label{#1}}
        \hfill
        \parbox[t]{#8\textwidth}{\caption {#7}\label{#5}}
    \end{center}
\end{figure} }

\def \myfiguress #1#2#3#4#5#6#7#8
{\begin{figure}[ht]
    \begin{center}
        \includegraphics[width=#2 \textwidth]{#1.eps}
        \hfill
        \includegraphics[width=#6 \textwidth]{#5.eps}
        \parbox[t]{#4\textwidth}{\caption {#3}\label{#1}}
        \hfill
        \parbox[t]{#8\textwidth}{\caption {#7}\label{#5}}
    \end{center}
\end{figure} }

\def \myf #1#2#3#4#5#6#7#8#9
{\begin{figure}[ht]
    \begin{center}
        \includegraphics[width=#2 \textwidth]{#1.eps}
        \hfill
        \includegraphics[width=#5 \textwidth]{#4.eps}
        \hfill
        \includegraphics[width=#8 \textwidth]{#7.eps}\\
        \parbox[t]{#2\textwidth}{\caption {#3}\label{#1}}
        \hfill
        \parbox[t]{#5\textwidth}{\caption {#6}\label{#4}}
        \hfill
        \parbox[t]{#8\textwidth}{\caption {#9}\label{#7}}
    \end{center}
\end{figure} }

\def \myff #1#2#3#4#5#6#7#8
{\begin{figure}[ht]
    \begin{center}
        \includegraphics[width=#2 \textwidth]{#1.eps}
        \hfill
        \includegraphics[width=#4 \textwidth]{#3.eps}
        \hfill
        \includegraphics[width=#6 \textwidth]{#5.eps}\\
        \parbox[t]{#7\textwidth}{\caption {#8}}
     \end{center}
\end{figure} }

\def\myfigure #1#2#3#4
{\begin{figure}[ht]\begin{center}
\includegraphics[width=#2 \textwidth, angle=-90]{#1.eps}
\parbox[t]{#4\textwidth}{\caption{#3}\label{#1}}
\end{center}\end{figure}}

\def\myfigured #1#2#3#4
{\begin{figure}[ht]\begin{center}
\includegraphics[width=#2 \textwidth]{#1.eps}\\
\vskip 2 cm
\parbox[t]{#4\textwidth}{\caption{#3}\label{#1}}
\end{center}\end{figure}}

\date{\today}
\begin{document}
\title{Anisotropic cosmological models with spinor field and viscous fluid in
presence of a $\Lambda$ term: qualitative solutions}
\author{Bijan Saha and Victor Rikhvitsky}
\affiliation{Laboratory of Information Technologies\\
Joint Institute for Nuclear Research, Dubna\\
141980 Dubna, Moscow region, Russia} \email{bijan@jinr.ru}
\homepage{http://www.jinr.ru/~bijan/}

\begin{abstract}

The study of a self-consistent system of nonlinear spinor and
Bianchi type I gravitational fields in presence of a viscous fluid
and $\Lambda$ term with the spinor field nonlinearity being some
arbitrary functions of the invariants $I$ an $J$ constructed from
bilinear spinor forms $S$ and $P$, generates a multi-parametric
system of ordinary differential equations \cite{saharrp,grqcnlsp}.
A qualitative analysis of the system in question has been thoroughly
carried out. A complete qualitative classification of the mode of
evolution of the universe given by the corresponding dynamic system
has been illustrated.
\end{abstract}

\keywords{Spinor field, Bianchi type I (BI) model, Cosmological constant,
viscous fluid, qualitative analysis}

\pacs{03.65.Pm and 04.20.Ha}

\maketitle

\bigskip


\section{Introduction}

Though the investigation of relativistic cosmological models
usually has the energy momentum tensor of matter generated by a
perfect fluid, to consider more realistic models one must take
into account the viscosity mechanisms
\cite{mis1,mis2,wein,murphy,belin}. On the other hand in the
recent years anisotropic cosmological models with nonlinear spinor
field have been extensively studied due to the facts that (i)
introduction of nonlinear spinor field into the system leads to
the isotropization of initially anisotropic universe
\cite{sahajmp,sahagrg,sahaprd}; (ii) spinor field nonlinearity in
some cases can give rise to singularity-free solutions
\cite{sahaprd,sahaecaa}; (iii) spinor field can be considered as
one of the possible candidate to explain the late time
acceleration of the universe \cite{sahaprd06,kramer}.

Given the importance of viscous fluid and spinor field to model a
realistic universe, recently we have considered a self-consistent
system of nonlinear spinor field and a gravitational field
described by a Bianchi type I (BI) cosmological model filled with
viscous fluid \cite{saharrp,grqcnlsp}. In \cite{saharrp,grqcnlsp}
we have thoroughly studied the corresponding field equations.
Exact solutions of the field equations were given in terms of
volume scale of the BI space-time $\tau$. Multi-parametric system
of equations for volume scale $\tau$, energy density of the
viscous fluid $\ve$ was solved for some special choice of bulk and
share viscosity. Given the richness of the system mentioned above
in this report we study it qualitatively for more general cases.
 In the absence of viscosity the system allows integrals of motion,
whereas it is either impossible obtain, or there is no first integral
at all in general when the viscosity is taken into account. As a result
the study of possible modes of evolutions becomes very difficult.
Undoubtedly, the result of the investigation should be presented in the
form of numerical values and at the same time can not be reduced to a
representation of some causally found examples of numerical solutions.
Clearly, it should be a classification of modes of the evolution in the
parametric space. Actually, it is the task of qualitative analysis.

\section{Basic equations}

We consider a self-consistent system of nonlinear spinor and
Bianchi type-I (BI) gravitational fields filled with a viscous
fluid in presence of a cosmological term. As it was shown in
\cite{saharrp,grqcnlsp}, the components of the spinor field and
metric functions can be expressed in terms of volume scale $\tau$
of the BI space-time. So one needs to find the function $\tau$,
explicitly. Corresponding equation can be derived from Einstein
equations and Bianchi identity [a detailed description of this
procedure can be found in \cite{saharrp,grqcnlsp}]. For
convenience, we also define the generalized Hubble constant. The
system then reads:

\begin{subequations}
\label{HVen1}
\begin{eqnarray}
\dot \tau &=& 3 H \tau, \label{tauH}\\
\dot {H} &=& \frac{1}{2}\bigl(3 \xi H - \omega\bigr) - \bigl(3 H^2
- \kappa\ve  - \Lambda\bigr) + \frac{\kappa}{2} \Bigl(
\frac{m}{\tau} + \frac{\lambda (n-2)}{\tau^{n}} \Bigr),
\label{Hn}\\
\dot {\ve} &=& 3 H\bigl(3 \xi H - \omega\bigr) + 4 \eta \bigl(3
H^2 - \kappa \ve  - \Lambda\bigr) - 4 \eta \kappa\Bigl[
\frac{m}{\tau} - \frac{\lambda}{\tau^n}\Bigr]. \label{Ven}
\end{eqnarray}
\end{subequations}
Here $\kappa$ is the Einstein's gravitational constant, $\Lambda$
is the cosmological constant, $\lambda $ is the self-coupling
constant, $m$ is the spinor mass and $n$ is the power of
nonlinearity of the spinor field (here we consider only power law
nonlinearity). In \eqref{HVen1} $\eta$ and $\xi$ are the bulk and
shear viscosity, respectively and they are both positively
definite, i.e.,
\begin{equation}
\eta > 0, \quad \xi > 0.
\end{equation}
They may be either constant or function of time or energy. We
consider the case when
\begin{equation}
\eta = A \ve^{\alpha}, \quad \xi = B \ve^{\beta}, \label{etaxi}
\end{equation}
with $A$ and $B$ being some positive quantities. For $p$ we set as
in perfect fluid,
\begin{equation}
p = \zeta \ve, \quad \zeta \in (0, 1]. \label{pzeta}
\end{equation}
Note that in this case $\zeta \ne 0$, since for dust pressure,
hence temperature is zero, that results in vanishing viscosity.
Note that a system in absence of spinor field has been studied in
\cite{Bmpla,physD}. In that case the corresponding system is
analogical to the one given in \eqref{HVen1} without the third
terms in \eqref{Hn} and \eqref{Ven}.

\section{Qualitative analysis}

Research on the behavior of the dynamic system given by a system of
ordinary differential equations implies the survey of all possible
scenarios of development for different values of the problem
parameters. It is necessary to understand at least how the process
of evolution comes to an end if it does so at infinitively large
time for a given set of initial conditions which can be given
anywhere.

So, under the specific behavior of the system we understand the
phase portrait of the system, i.e., the family of integral curves,
covering the total phase space. It is easy to imagine as far as
any point of the space can be declared as the initial one and at
least one integral curve will pass through it (or it will be fixed
point).

Certainly, it is difficult to imagine such a set of curves. In
many cases, close (and not only) curves transform into each other
at some diffeomorphism of space. These curves are known as
topologically equivalent. The differences between them are not
very important for our study. They all behave in the same manner.
This relation - "the relation of equivalence" - divides the family
of curves into the classes of equivalence. For graphical
demonstration it will be convenient to present at least one
representative of each class.

The change of the value of problem parameters not always results
in significant change of the phase portrait. Repeating this method, we
say that one family of integral curves (covering the total space)
for the given set of parameters is equivalent to the other for
another set of parameters, if there exists a diffeomorphism of
space transforming the first family into the second. It is clear
that there occurs the division into the classes of equivalence, and
we are not very interested in differences between equivalent
families. We argue that the corresponding changes in parameters
do not alter anything on principle. So it is sufficient to
demonstrate only one phase portrait for a given set of parameters
underlining the features of the given class.

However, for some critical relations between the parameters there
occurs significant changes. These are the boundary relations of
parameters, dividing, as usual, parameter space into regions of
similar behavior. Thus accomplishes the qualitative classification
of the mode of evolution of dynamic system. Now, giving the
concrete value of parameters, we can define which region of
parameters they correspond to, thus define the type of behavior.
Moreover, given the specific initial conditions, we can answer
the question to which region of phase space the evolution of the
system lead in time.

In our cosmological model, numerical parameters $A$, $\alpha$, $B$,
$\beta$ are related to the viscosity, while $\lambda$ and
$\Lambda$ are the (self)-coupling and cosmological constants.

Initially, we consider the system of Einstein and Dirac equations.
Solving these equations, we find the components of the spinor
field and metric functions $a,\,b,\,c$ in terms of volume scale
$\tau = abc$ of the BI universe. Finally, in order to find $\tau$
from Einstein equations and Bianchi identity, we deduce three
first order ordinary differential equations. Further for
convenience we introduce a new function $\nu$ inverse to $\tau$,
i.e., $\nu = 1/\tau$.

The fact that the system has the dimension greater than 2,
strongly complicates qualitative analysis. Note that well known
Lorentz system of three ordinary differential equations with
polynomial right hand side with degree less or equal to 2,
possesses in some region of parameter space chaotic behavior known
as a strange attractor and in that region there do not exist first
integrals (i.e., globally defined invariants). Though the set of
singularities is very simple, there exist only three singular
(fixed) points: two focus and one saddle. The presence of such
example does not allow us to make an optimistic conclusion on the
basis of simple construction of our system (with polynomials in
the right hand side and absence of singular points the in region
of space we are interest in, which is even dynamically closed.

Nevertheless, on the boundary of the the space $\epsilon=0$, as
well as $\nu=0$ ($\tau=+\infty$), which are dynamically closed
themselves, the complete classification has been done. The
dynamical closeness of these planes simultaneously as an obstacle
for penetration from positive octant $\epsilon>0$ $\land$ $\nu>0$
to the region with negative values. But, there are no
singularities, fixed points (there are fixed points on the
boundary) in the positive octant, we were not able to prove the
simplicity of its behavior, e.g., presence of first integrals, as
well as their absence.

Thus let us go back to the system \eqref{HVen1} in details. As it
was already mentioned, tt is convenient to define a new function
$\nu = 1/\tau$. In this case the obvious singularity that occurs
at $\tau = 0$ vanishes and $\nu = 0$ corresponds to $\tau =
\infty$ while $\nu = \infty$ to $\tau = 0$. The system
\eqref{HVen1} on account of \eqref{etaxi} takes the form:
\begin{subequations}
\label{HVen2}
\begin{eqnarray}
\dot \nu &=& - 3 H \nu, \label{tauH2}\\
\dot {H} &=& \frac{1}{2}\bigl(3 B \ve^\beta H - (1+\zeta)\ve
\bigr) - \bigl(3 H^2 - \ve - \Lambda \bigr) + \frac{1}{2}
\Bigl(m\nu + \lambda (n-2)\nu^{n-1} \Bigr),
\label{Hn2}\\
\dot {\ve} &=& 3 H\bigl(3 B \ve^\beta H - (1+\zeta)\ve \bigr) + 4
A \ve^\alpha \bigl(3 H^2 - \ve - \Lambda\bigr) - 4 A \ve^\alpha
\Bigl[ m\nu - \lambda\nu^n \Bigr]. \label{Ven2}
\end{eqnarray}
\end{subequations}
Let us now study the foregoing system of equations in details.

\subsection{Behavior of the solutions on $\nu = 0$ plane}

Let us first study the behavior of the functions $H$ and $\tau$ on
$\nu = 0$ plane. The plane $\nu = 0$ is dynamically invariant,
since $\dot \nu\bigl|_{\nu = 0} = 0.$ It should be emphasized that
the system in this case coincides with one in absence on spinor
field and was thoroughly studied in \cite{physD}. Nevertheless, we
write the results obtained in detail. In doing so we rewrite the
Eqs. \eqref{Hn2} and \eqref{Ven2} in the matrix form:
\begin{equation}
\left(
\begin{array}{c}
\dot {H}\\
\dot{\ve}
\end{array}
\right) = \left( \begin{array}{ccc} \kappa/2& & -1\\
3 H &  &4 \eta\end{array}\right) \left(
\begin{array}{c}
3 B \ve^\beta H - (1+\zeta)\ve\\
3 H^2 - \kappa \ve - \Lambda
\end{array}
\right). \label{eqmatr}
\end{equation}

{\bf a)} By virtu of linear independence of the columns of the
matrix of the Eq. \eqref{eqmatr} the critical points are the
solutions of the equations
\begin{subequations}
\label{cheq}
\begin{eqnarray}
3 B \ve^\beta H - (1+\zeta)\ve &=& 0, \label{ch1}\\
3 H^2 - \kappa \ve - \Lambda &=& 0. \label{ch2}
\end{eqnarray}
\end{subequations}
i.e., they necessarily lie on the parabola \eqref{ch2}. In view of
$$H = \frac{1 + \zeta}{3 B} \ve^{1-\beta}$$ which follows from
\eqref{ch1}, the Eq. \eqref{ch2} can be written as
\begin{equation}
 3 \kappa B^2 \ve^{1 + 2 \beta} - (1 + \zeta)^2 \ve^2 + 3
\Lambda B^2 \ve^{2 \beta} = 0. \label{ch1n}
\end{equation}
The solutions to the system \eqref{cheq} will be the roots of the
Eq. \eqref{ch1n}. The quantity of the positive roots of Eq.
\eqref{ch1n} according to Cartesian law is equal to the number of
changes of sign of the coefficients of equations or less than that
by an even number. So, for $\Lambda > 0$ and $ 1/2 < \beta < 1$ or
$\Lambda < 0$ and $\beta < 1/2$ the number of roots is either $2$
or zero. For the remaining cases, i.e., $ \Lambda > 0$ and $\beta
> 1$ or $\Lambda > 0$ and $\beta < 1/2$ or $\Lambda < 0$
and $\beta > 1/2$ there exists only one root.

In Table 1 classification of qualitatively different types of
evolution (phase portrait) depending on the parameters $\beta$,
$\Lambda$ and $(1+\zeta)/B$ are illustrated. The Fig. a) in Table
1 corresponds to the two types, namely $\beta < 1/2$ or $\beta =
1/2$ and $(1+\zeta)/B < \sqrt{3\kappa}$, as well as the Fig. i) to
the cases $\beta > 1$ or $\beta = 1$ and $(1+\zeta)/B <
\sqrt{3\kappa}$. The Figs. g) and h) of Table 1 cover all the four
cases for $\beta > 1/2$. The Figs. d) and e) contain pairs of
graphics (case with 0 or 2 singular points; case with 1 singular
point is also allowable since it possesses the frequency 2, i.e.,
two singular points merges to one saddle-knot and no other
qualitative change takes place). They covers 3 cases: d)
corresponds to $\beta \le 1/2$; e) corresponds to $\beta = 1/2$
and $(1+\zeta)/B < \sqrt{3\kappa}$ or $1/2 \le \beta \le 1$ and
$\beta = 1/2$ and  and $(1+\zeta)/B > \sqrt{3\kappa}$.

\begin{center}
\begin{tabular}{|c|c|c|c|c|}
  \hline
    &  & $\Lambda>0$ & $\Lambda=0$ & $\Lambda<0$ \\
  \hline
   $\beta<1/2$ & &
   \begin{tabular}{c} \includegraphics[width=0.18 \textwidth]{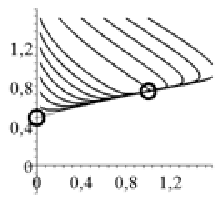} \\ a) \end{tabular} &
   \begin{tabular}{c} \includegraphics[width=0.18 \textwidth]{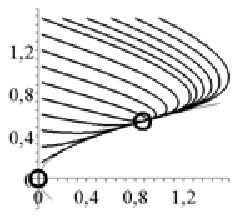} \\ b) \end{tabular} &
   \begin{tabular}{c} \includegraphics[width=0.18 \textwidth]{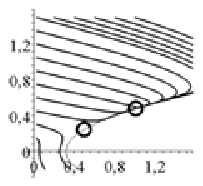} \\    \end{tabular} \\
  \cline{1-2} \cline{4-4}
   $\beta=1/2$
   & $\frac{1+\zeta}{B}> \sqrt{3\kappa}$ &
   &
   \begin{tabular}{c} \includegraphics[width=0.18 \textwidth]{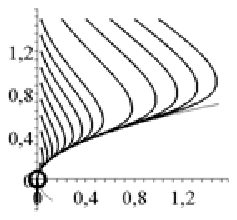} \\ c) \end{tabular} &
   \begin{tabular}{c} \includegraphics[width=0.18 \textwidth]{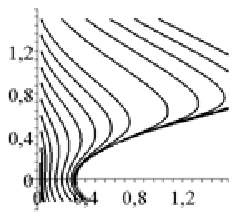} \\ d) \end{tabular} \\
   \cline{2-4}
   & $\frac{1+\zeta}{B}< \sqrt{3\kappa}$ &
   \begin{tabular}{c} \includegraphics[width=0.18 \textwidth]{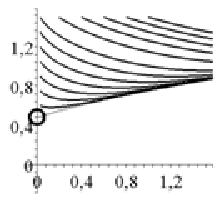} \\  \end{tabular} &
   \begin{tabular}{c} \includegraphics[width=0.18 \textwidth]{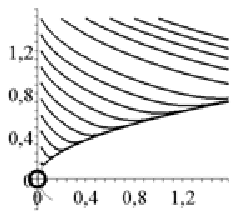} \\ f) \end{tabular} &
   \\
  \cline{1-2} \cline{4-5}
   $1/2<\beta<1$ & &
   \begin{tabular}{c} \includegraphics[width=0.18 \textwidth]{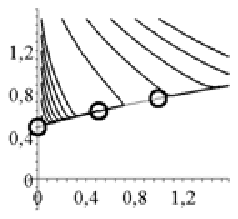} \\ e) \end{tabular} &
   \begin{tabular}{c} \includegraphics[width=0.18 \textwidth]{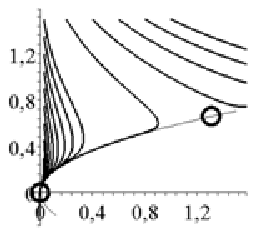} \\ g) \end{tabular} &
   \begin{tabular}{c} \includegraphics[width=0.18 \textwidth]{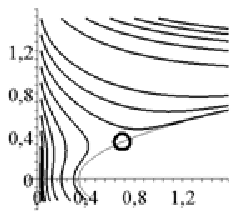} \\ h) \end{tabular} \\
  \cline{1-2}
   $\beta=1$
   & \begin{tabular}{c} \ \\ $\frac{1+\zeta}{B}>3\Lambda$ \\ \ \end{tabular} &
   &
   &
   \\
  \cline{2-3}
   & $\frac{1+\zeta}{B}<3\Lambda$ &
   \begin{tabular}{c} \includegraphics[width=0.18 \textwidth]{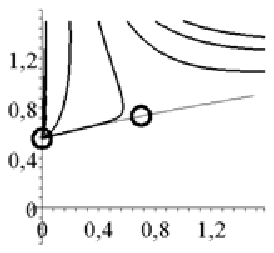} \\ i) \end{tabular} &
   &
   \\
  \cline{1-2}
   \begin{tabular}{c} \  $\beta>1$  \ \end{tabular}
    & &   &    &
   \\
  \hline
 \end{tabular}
\vskip 1 cm {\bf Table 1.} Classification of qualitatively
different types of evolution (phase portrait) depending on the
parameters $\beta$, $\Lambda$ and $(1+\zeta)/B$
\end{center}

Since, the equation for $\ve$ only contains $\eta$, the energy
density for nontrivial $\eta$ undergoes essential changes, whereas
$H$ and $\tau$ remain virtually unchanged.

The types of critical points lying on the integral curve
alternate: $\ldots$ saddle, attracting knot, saddle $\ldots$. So
it is sufficient to consider the case with maximum number of
roots. Taking into account the Eqs. \eqref{Ven} and \eqref{ch2}
let us now calculate
\begin{eqnarray}
\lim_{\ve \to +\infty} \frac{\dot \ve}{3 H \ve} &=& \lim_{\ve \to
+\infty} \frac{\sqrt{3} B \ve^\beta \sqrt{\kappa \ve + \Lambda} -
\ve (1 + \zeta)}{\ve} \nonumber \\ &=& \sqrt{3} B \sqrt{\kappa
\ve^{(2\beta -1)} + \Lambda \ve^{-2}} - (1 + \zeta) =
\left\{\begin{array}{ccccc} -(1 + \zeta) < 0,& \beta < 1/2, \\
B\sqrt{3\kappa} - (1 + \zeta), & \beta = 1/2, \\
+ \infty  >  0, & \beta > 1/2. \\
\end{array}\right.
\end{eqnarray}
So, the latest critical point for $\beta < 1/2$ is attracting knot
and for $\beta > 1/2$ is saddle. In case of $\beta = 1/2$ we have
either saddle if $B\sqrt{3\kappa} - (1 + \zeta) > 0$ and
attracting knot otherwise.

{\bf b)}  It is obvious that if $\,\,\,\Lambda \geq 0\,\,\,$ the
points of intersection of the boundary are the critical points
\begin{subequations}
\begin{eqnarray}
H &=& \pm \sqrt{\Lambda/3},\\
\ve &=& 0.
\end{eqnarray}
\end{subequations}

{\bf c)} For $H < 0$ there may exist critical points , if the
columns of the matrix of \eqref{eqmatr} are linearly dependent. In
that case the critical points are the roots of the equation
\begin{equation}
3 \kappa (\zeta -1)\ve + 6 \kappa^2 A B \ve^{\alpha + \beta}+ 8
\kappa^2 A^2 \ve^{2 \alpha} - 6 \Lambda  = 0,
\end{equation}
and
\begin{equation}
H = -\frac{2}{3} \kappa A \ve^{\alpha}.
\end{equation}

In case of $\eta = 0$ the roots of the characteristic equation
\begin{equation}
\Bigl|\frac{D({\dot H},\,{\dot \ve})}{D(H,\,\ve)} - \mu \Bigr| =
0,
\end{equation}
are
\begin{equation}
\mu_{1,2} = \frac{3 \kappa \xi \pm \sqrt{9 \kappa^2 \xi^2 + 48
\Lambda (1 + \zeta)}}{4}.
\end{equation}
The critical point $\,\,\,(H,\,\ve) = (0,\,2\Lambda/[\kappa(\zeta
- 1)])\,\,\,$ is of type divergent focus if $\,\,\Lambda > - 9
\kappa^2 \xi^2/[48(1 + \zeta)]\,\,$ or divergent knot if
$\,\,\Lambda < - 9 \kappa^2 \xi^2/[48(1 + \zeta)]\,\,$.

\subsubsection{Integral curves}

For $\Lambda \ge 0$ the solutions starting from the upper
half-plane $H > 0$ cannot enter into the lower one. For $\Lambda <
0$ some of the solutions may enter into the lower half-plane
through the segment $H = 0$ and $\Lambda \leq 0 \leq \ve$ and
never returns back, since $\dot{H}|_{H=0} < 0.$

\subsection{Behavior of the solutions on $\ve = 0$ plane}

The plane $\ve = 0$ is dynamic invariant, since $\dot
\ve\bigl|_{\ve = 0} = 0$. Depending on the sign of $H$ this plane
is either attractive or repulsive, namely, for $H > 0$ it is
attractive and for $H < 0$ it is repulsive, since
$$\frac{\partial \dot \ve}{\partial \ve} = -3H(1 + \zeta) < 0.$$
On $\ve = 0$ plane the system \eqref{HVen2} takes the form

\begin{subequations}
\label{nu=0}
\begin{eqnarray}
\dot \nu &=& -3H\nu, \label{nuve0}\\
\dot H &=& -3H^2 + \Lambda + \frac{1}{2}(m\nu +
\lambda(n-2)\nu^{n-1}). \label{Hve0}
\end{eqnarray}
\end{subequations}
The system \eqref{nu=0} has the following integrals:
\begin{subequations}
\label{nu=01int}
\begin{eqnarray}
3H^2 &=& C \nu^2 + m \nu + \Lambda - \frac{\lambda (n-2)\nu^{n-1}}{n-3}, \quad n > 3, \label{int1}\\
3H^2 &=& C \nu^2 + m \nu + \Lambda - \nu^2 \ln(\nu), \quad  \quad \quad \,\,\, n = 3, \label{int2}\\
3H^2 &=& C \nu^2 + m \nu + \Lambda, \qquad \qquad \qquad \quad
\,\,\,\,\,\, n = 2, \label{int3}
\end{eqnarray}
\end{subequations}
where $C$ is some arbitrary constant.

The characteristic equation of nontrivial singular points on $\ve
= 0$ plane for the system \eqref{HVen1} takes the form
\begin{equation}
\lambda (n-2) \nu^{n-1} + m \nu + 2 \Lambda = 0.
\end{equation}
Depending on changes of signs in the sequence of $\lambda$, $m$,
$\Lambda$ it has one, two or no solutions.

In Table 2 we illustrated the phase-portrait on $\ve = 0$ plane
for a positive and a negative $\Lambda$, respectively for $n=3$.

\vspace{1cm}
\begin{center}
\begin{tabular}{|c|c|c|}
  \hline
      $\Lambda<0$ &  $\Lambda>0$ \\
  \hline
   \begin{tabular}{c} \includegraphics[width=0.30 \textwidth]{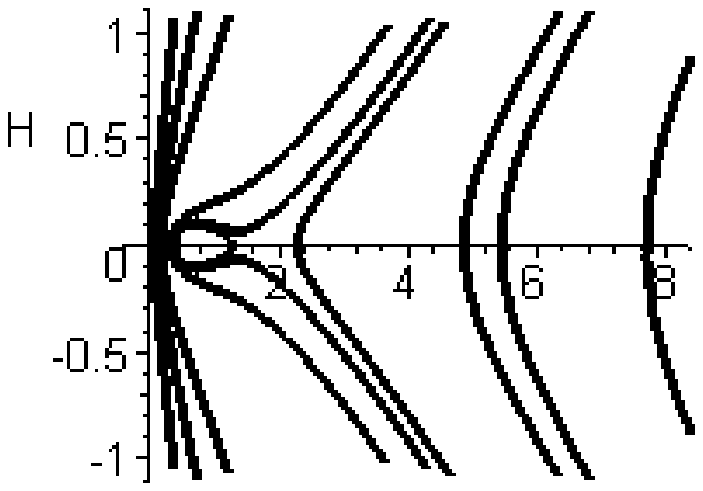}\\a) \\ \end{tabular} &
   \begin{tabular}{c} \includegraphics[width=0.30 \textwidth]{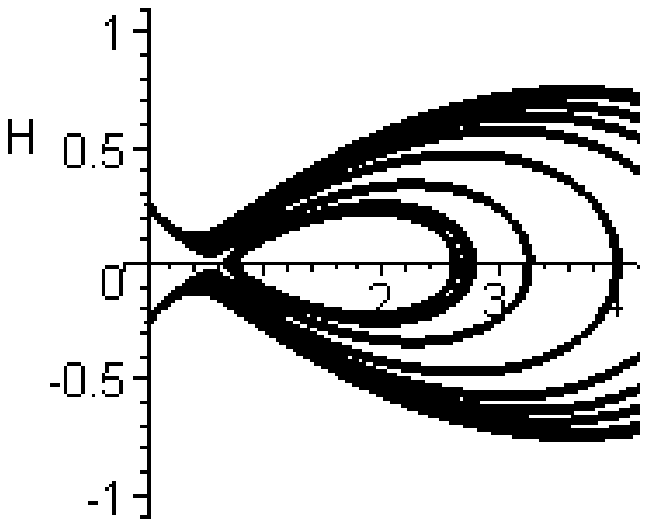}\\b) \\ \end{tabular} \\
  \hline
\end{tabular}
\vskip 1 cm {\bf Table 2.} Classification of qualitatively
different types of evolution (phase portrait) on $\ve = 0$ plane
for $n = 3$
\end{center}

In Table 3 we have graphically illustrated the $H-\nu$ phase
portrait on the $\ve=  0$ plane for different $\Lambda$. As it was
mentioned earlier, here we deal with the multi-parametric system
of ordinary nonlinear differential equation. In doing so we
consider all possible variants independent to their physical
validity. Therefore, we demonstrate the results obtained for a
negative spinor mass ($m < 0$).

\begin{center}
\begin{tabular}{|c|c|c|c|c|}
  \hline
     & $\Lambda<0$ & $\Lambda=0$ & $\Lambda>0$ \\
  \hline
   $m<0$ &
   \begin{tabular}{c} \includegraphics[width=0.25 \textwidth]{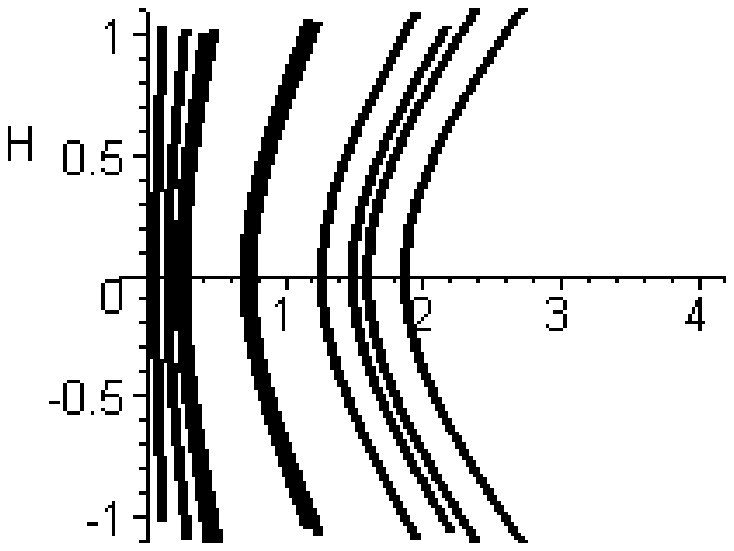}\\a) \\ \end{tabular} &
   \begin{tabular}{c} \includegraphics[width=0.25 \textwidth]{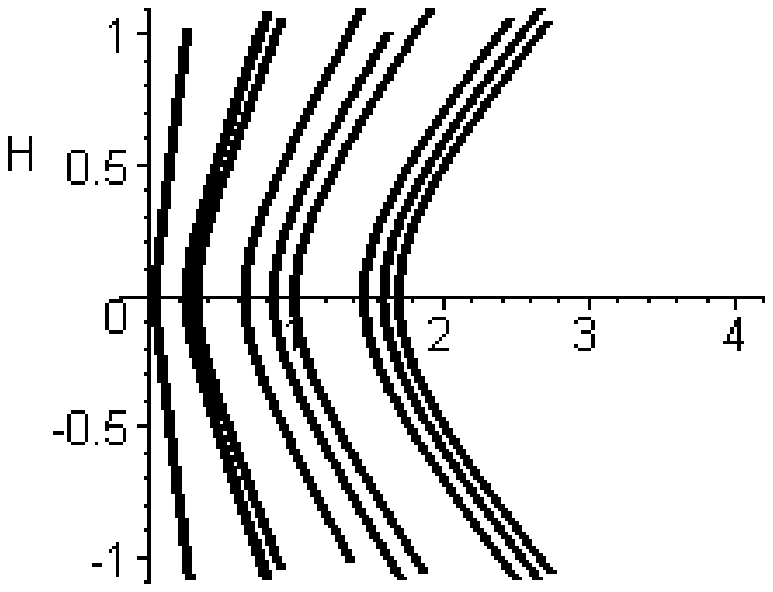}\\b) \\ \end{tabular} &
   \begin{tabular}{c} \includegraphics[width=0.25 \textwidth]{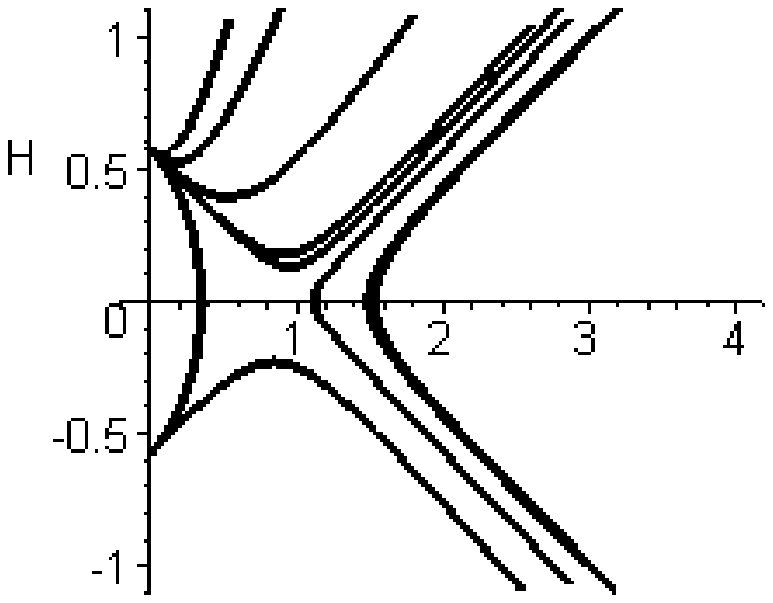}\\c) \\ \end{tabular} \\
  \hline
   $m=0$ &
   \begin{tabular}{c} \includegraphics[width=0.25 \textwidth]{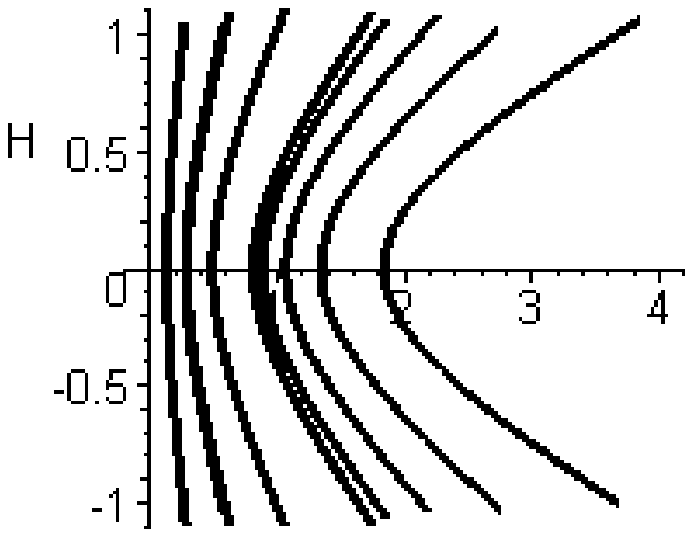}\\d) \\ \end{tabular} &
   \begin{tabular}{c} \includegraphics[width=0.25 \textwidth]{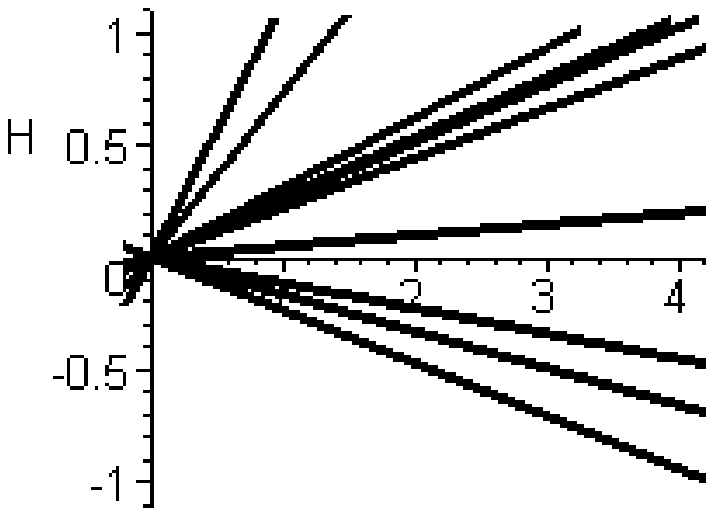}\\e) \\ \end{tabular} &
   \begin{tabular}{c} \includegraphics[width=0.25 \textwidth]{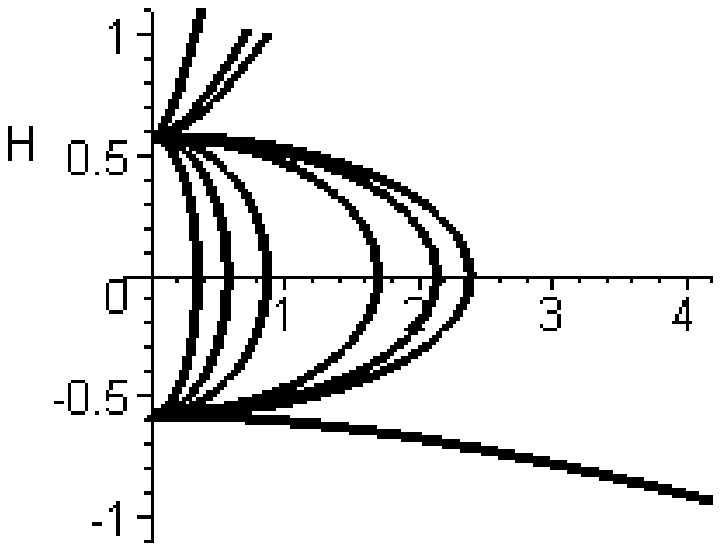}\\f) \\ \end{tabular} \\
  \hline
   $m>0$ &
   \begin{tabular}{c} \includegraphics[width=0.25 \textwidth]{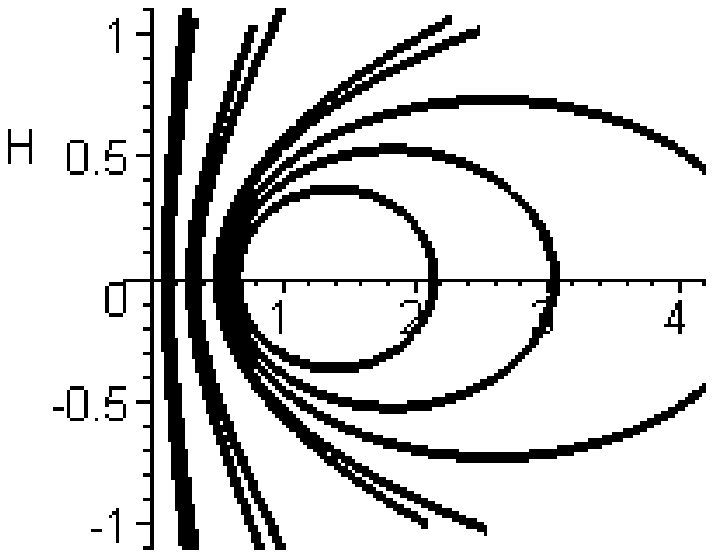}\\g) \\ \end{tabular} &
   \begin{tabular}{c} \includegraphics[width=0.25 \textwidth]{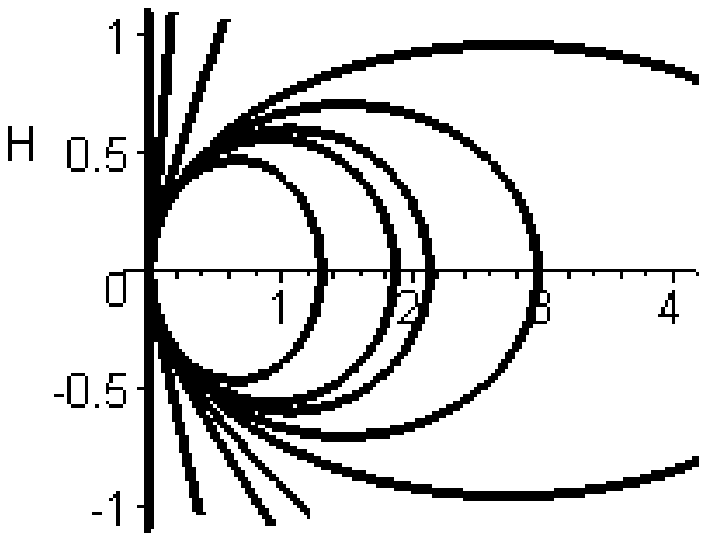}\\h) \\ \end{tabular} &
   \begin{tabular}{c} \includegraphics[width=0.25 \textwidth]{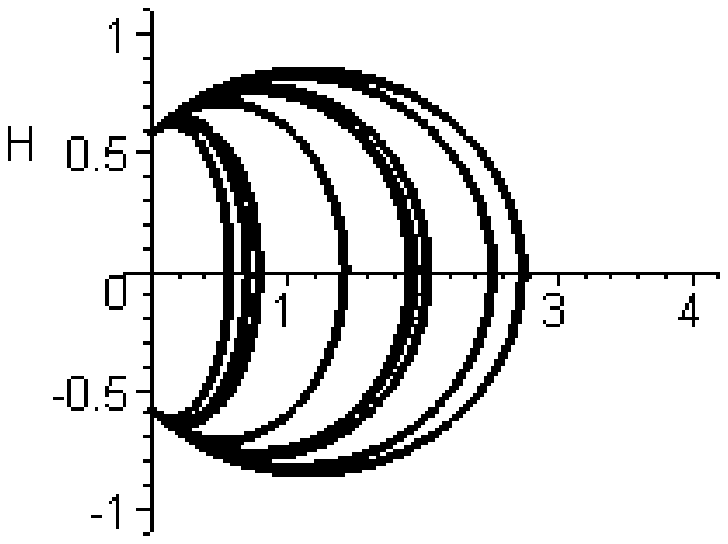}\\i) \\ \end{tabular} \\
  \hline
\end{tabular}
\vskip 1 cm {\bf Table 3.} Classification of qualitatively
different types of evolution (phase portrait) on $\ve = 0$ plane
for $n = 2$
\end{center}

If the right hand side of \eqref{nu=01int} possesses two positive
roots with $H$ being positive between them, then on the plane $\ve
= 0$ there occurs closed cycle. It is obvious that there can be no
more than three roots, hence there cannot be non-concentric
cycles. As a result, near the plane $\ve = 0$ there might be
cyclic oscillations.

The singular point around which the oscillation takes place has $H
= 0$, and therefore, the trajectory of oscillation partially
passes in the region which is attractive to the plane $\ve = 0$
and partially in the region that is repulsive. In the long run in
the repulsive region at some moment the growth of $\ve$ becomes
dominant. It results in the fact that $\ve$ becomes infinity
within a finite range of time.

\subsubsection{Invariants of evolution}

The system \eqref{HVen2} in absence of viscosity, i.e., under
$\eta = 0$ and $\xi = 0$ possesses the following first integrals
\begin{subequations}
\label{F}
\begin{eqnarray}
F_1 &=& \frac{\ve}{\nu^{1+\zeta}}, \label{F1}\\
F_2 &=&
\frac{(H^2-\ve-\Lambda-m\nu)}{\nu^2}+\lambda\frac{n-2}{n-3}\nu^{n-3}.
\label{F2}
\end{eqnarray}
\end{subequations}

The first of them \eqref{F1} remains to be the first integral even
after the introduction of bulk viscosity $\xi$. The second one,
i.e., Eq. \eqref{F2} under $\xi \ne 0$ ceases to be the integral
of motion. Nevertheless, the introduction of bulk viscosity during
the course of time generates definite displacement of the surface
given by the formula \eqref{F2}, which allows one qualitatively,
i.e., based only on the continuity, compile the representation
about the possible ways of evolution.

\subsection{Qualitative analysis of the complete system}

Harnessing the Tables 2, 3 as well as 1, helps one to understand
the 3D phase portrait leaning on the continuous dependence of the
velocity fields of the coordinates $\nu, H, \ve$ of phase space.

In order to cover the infinite phase space completely, it is
mapped on coordinate parallelepiped with its axes being the the
arc-tangent of the corresponding coordinates. The lower horizontal
plane always represents the $\ve = 0$ plane.

It should be noted that the introduction of spinor field notably
complicates the evolution of the system. Contrary to the system in
absence of the spinor field, the initial condition with $H < 0$
already does not prevent in many cases thanks to the evolution of
volume scale entering the half-space $H > 0$ and thereupon, from
the greater value of $H$ repeats the evolution, approaching to the
$\nu = 0$ plane and displaying the classification from the table
1. In the vicinity of the borders $\ve = 0$ and $\nu = 0$ the
integral curves closely repeats the integral curves on the sides,
each time at least to some extent.

The general property of all the cases is the fact that in the
half-space $H > 0$ the velocity vectors are directed to the $\ve =
0$ plane, while in the other half opposite to it. As a result all
he invariant curves fall on $\ve = 0$, though not necessarily
reach it.

In the Figs. \ref{tau1} - \ref{3d4} we have illustrated the the
function inverse proportional to volume scale, i.e., $\nu(t)$
[Figs. \ref{tau1},\ref{tau2},\ref{tau3},\ref{tau4}], volume scale
$\tau(t)$ [Figs. \ref{tau1a},\ref{tau2a},\ref{tau3a},\ref{tau4a}]
and the phase portrait in the $\nu,H,\ve$ space [Figs.
\ref{3d1},\ref{3d2},\ref{3d3},\ref{3d4}] , corresponding to the
positions c,g,h, and i of the Table 2. In doing so we used the
following values of the parameters: $\alpha = 4$, $\beta = 1$,
$\zeta = 0.5$, $A = 1$, $B = 1$ and $n = 4$.The positions h and i
correspond to the geometrically cyclic regime, but the case h
possesses fixed point on cyclic integral curve, hence corresponds
to the intermediate stage between periodic and non-periodic. The
positions c and g correspond to the non-periodic evolution.

The bold black line in the 3D figures [Figs.
\ref{3d1},\ref{3d2},\ref{3d3},\ref{3d4}] corresponds to the the
functions $\nu(t)$ and $\tau(t)$ presented in the preceding
Figures [Figs.
\ref{tau1},\ref{tau2},\ref{tau3},\ref{tau4},\ref{tau1a},\ref{tau2a},\ref{tau3a},\ref{tau4a}].
Figs. \ref{tau1},\ref{tau1a},\ref{3d1} and
\ref{tau4},\ref{tau4a},\ref{3d4} correspond to the case k of Table
1, Fig. \ref{tau2},\ref{tau2a},\ref{3d2} to i and Fig.
\ref{tau3},\ref{tau3a},\ref{3d3} to h, respectively.

The simplest case is illustrated in Figs.
\ref{tau4},\ref{tau4a},\ref{3d4}. The integral curves beginning
from anywhere falls into $\nu = 0$. The other peculiarities of
this curve are of no significance for the volume scale.

In case of Fig. \ref{tau1},\ref{tau1a},\ref{3d1}, there exists
hyperbolic singular point on the $\ve = 0$ plane, so additional to
the Fig. \ref{tau4},\ref{tau4a},\ref{3d4} there exists curves
tending to the side $\nu = \infty$.

The cyclic trajectories of Figs. \ref{tau2},\ref{tau2a},\ref{3d2}
beginning at the vicinity of $\ve = 0$ plane, gradually moving
away, approach to the $\ve = \infty$ plane, and the plane itself
in the long run turns out to be attracting one with the edges $H =
\pm \infty$.

The geometrically cyclic lines in Figs.
\ref{tau3},\ref{tau3a},\ref{3d3} in the $\ve = 0$ plane contains
fixed point, hence the behavior of the corresponding trajectory is
very similar to the one given in Figs.
\ref{tau4},\ref{tau4a},\ref{3d4}.


\myf{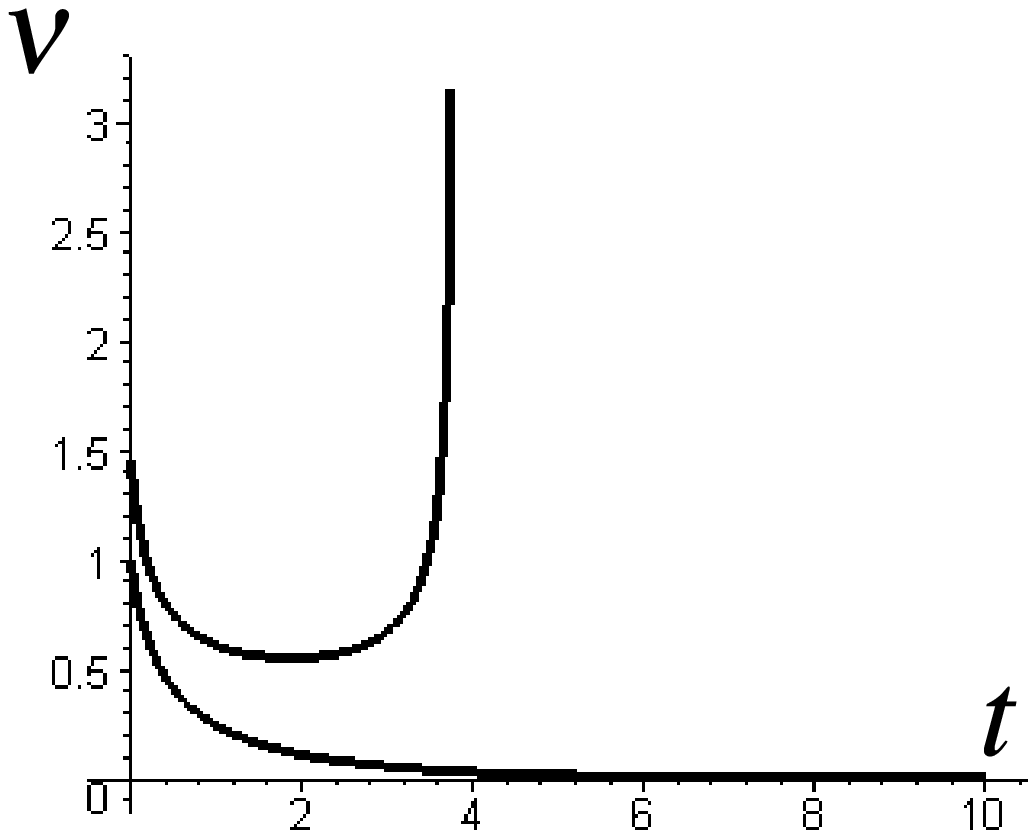}{0.25}{Evolution of function inverse to volume
scale}{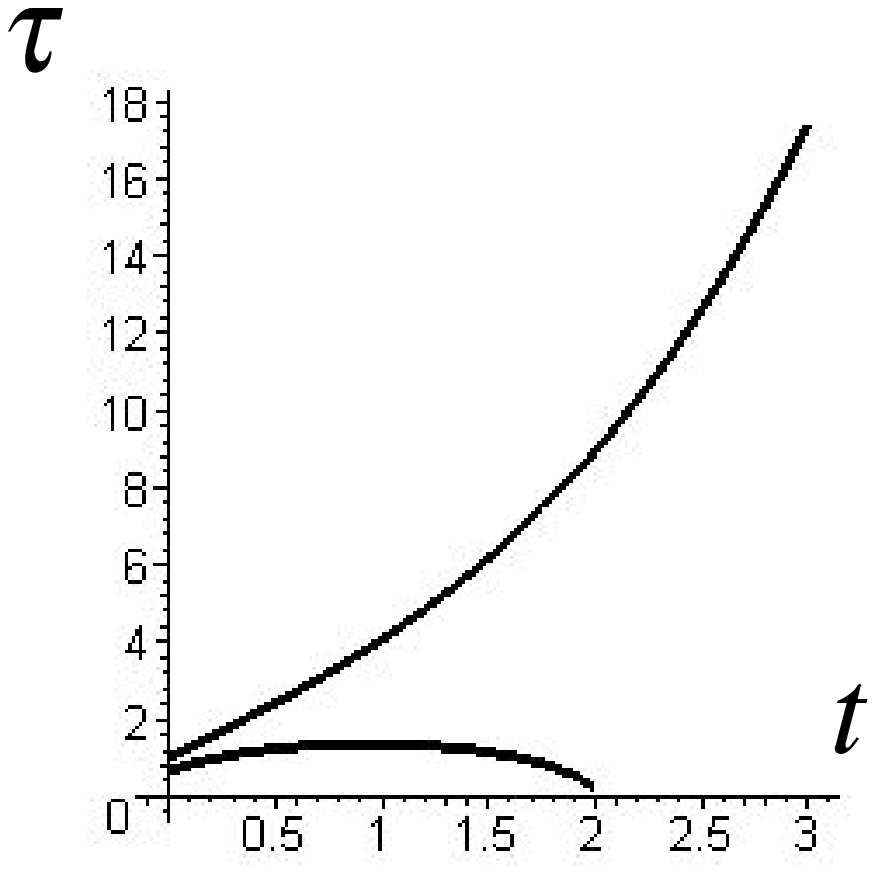}{0.20}{Evolution of volume scale}{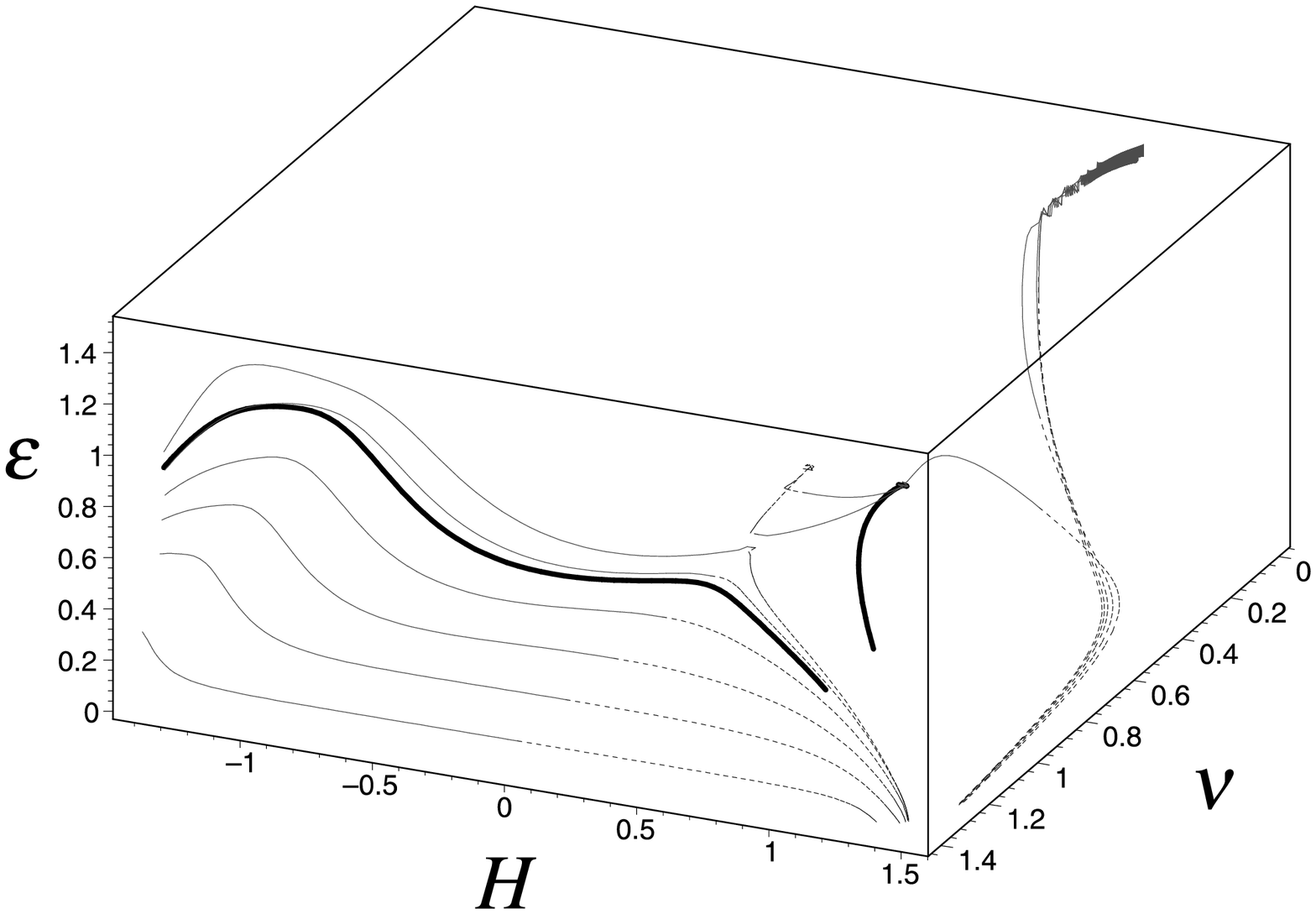}{0.20}{3D view
in $\nu,H,\ve$ space}

\myf{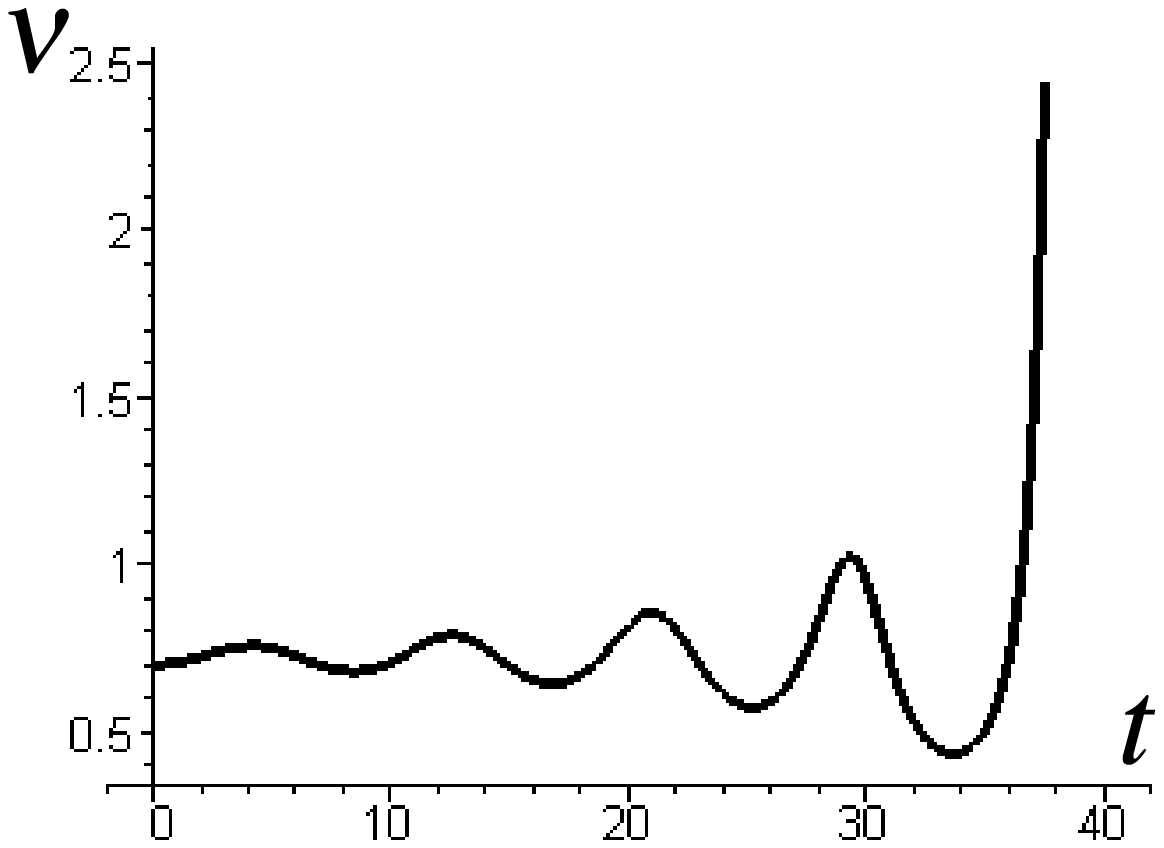}{0.25}{Evolution of function inverse to volume
scale}{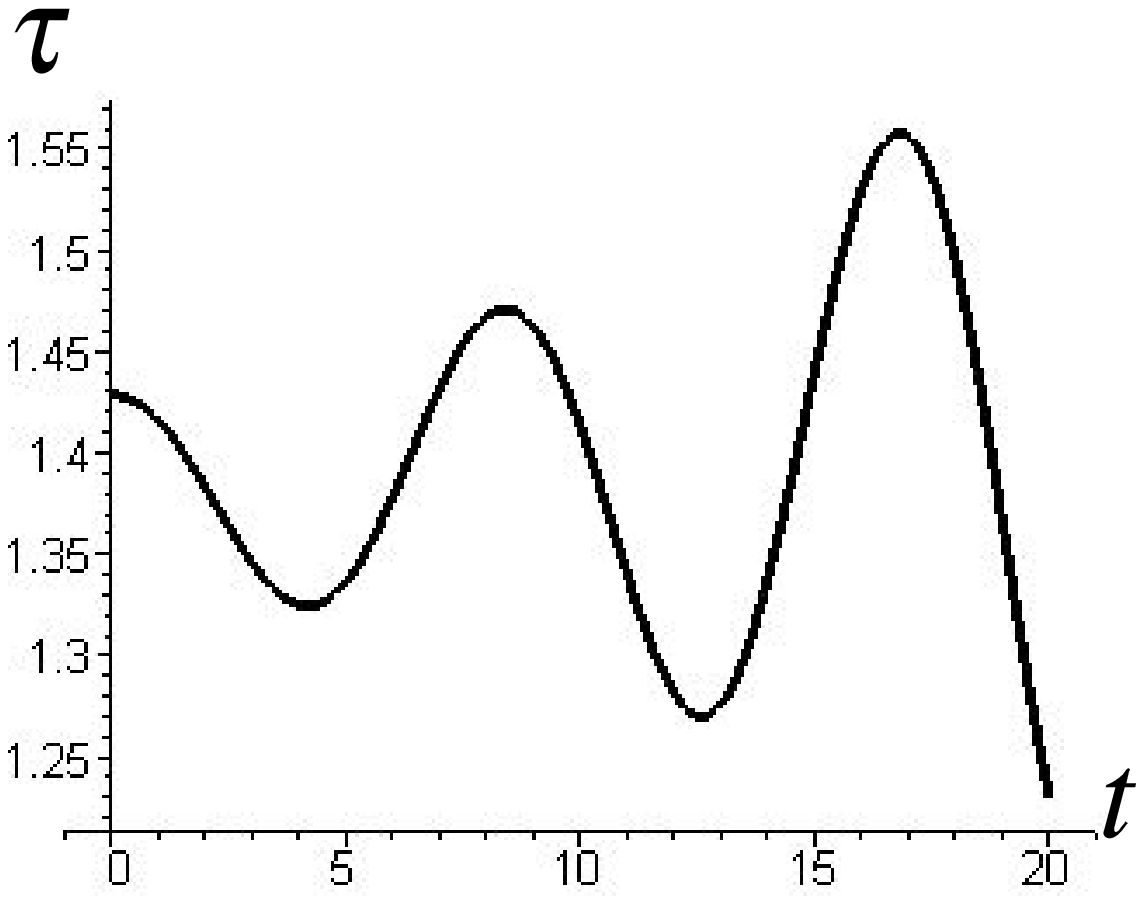}{0.20}{Evolution of volume scale}{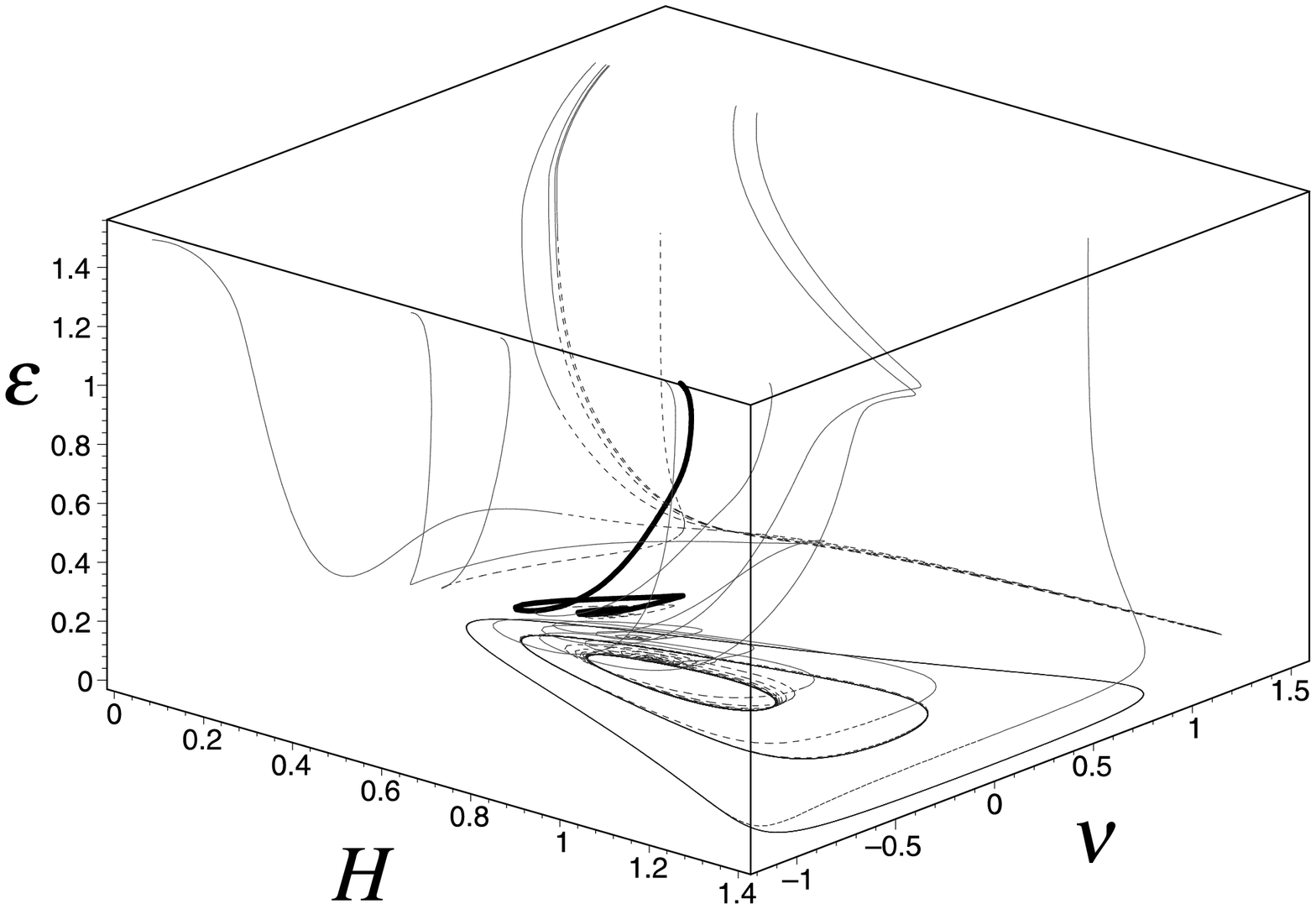}{0.20}{3D view
in $\nu,H,\ve$ space}

\myf{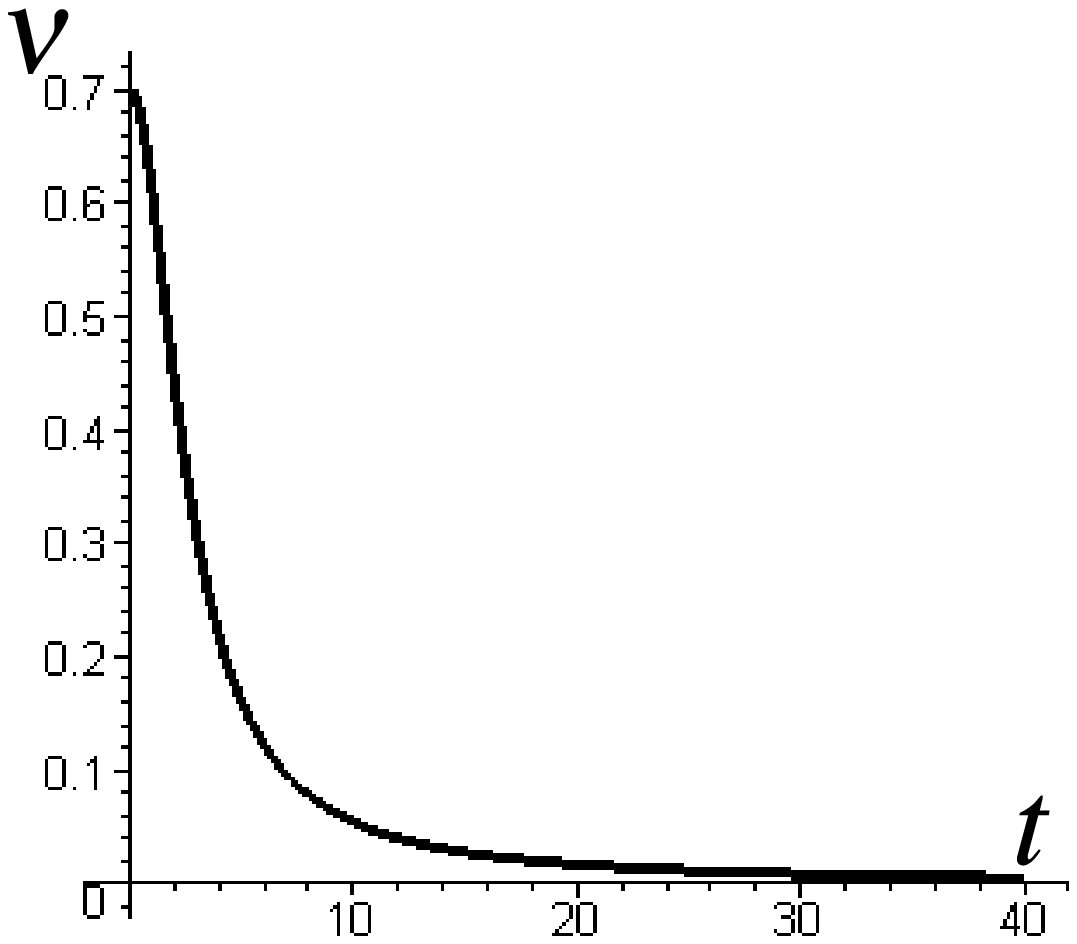}{0.25}{Evolution of function inverse to volume
scale}{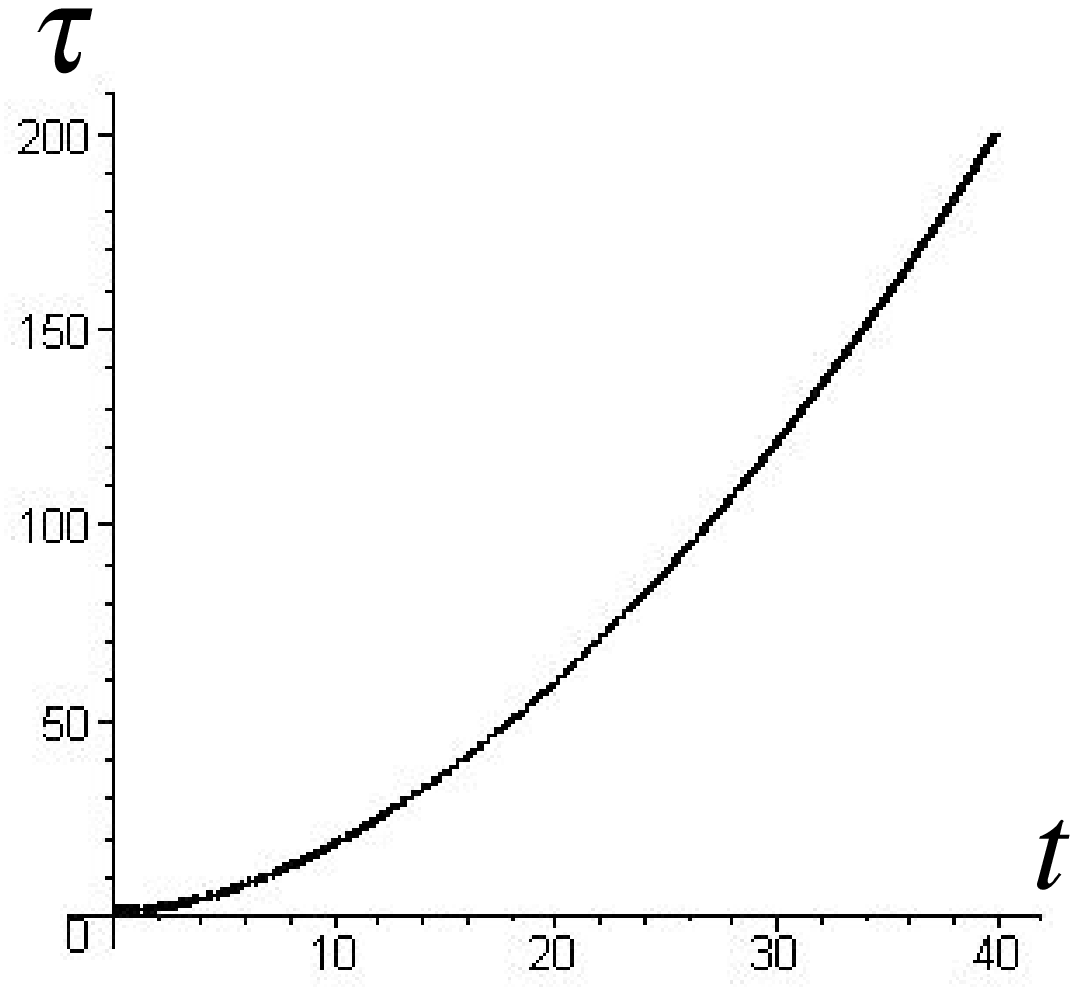}{0.20}{Evolution of volume scale}{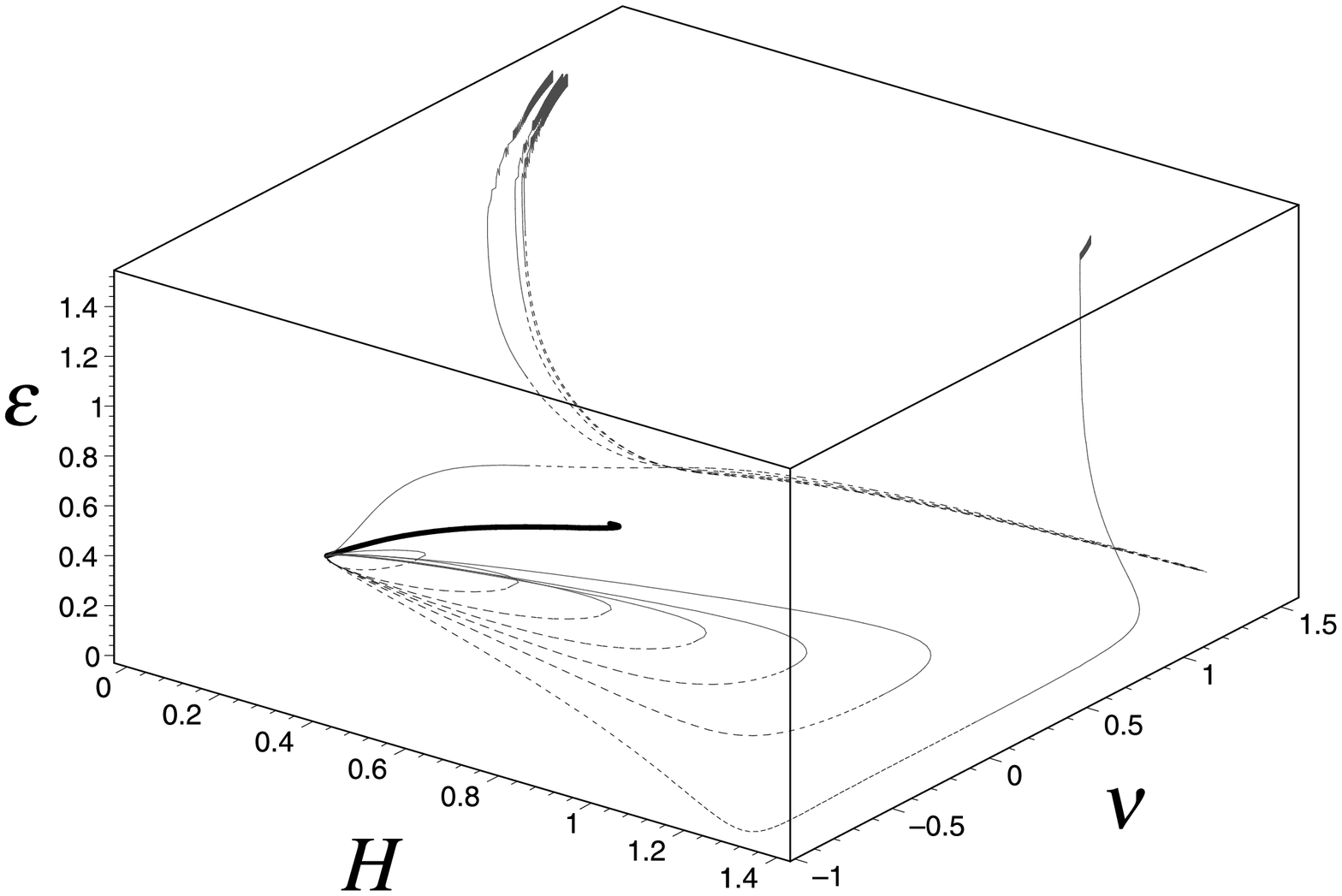}{0.20}{3D view
in $\nu,H,\ve$ space}

\myf{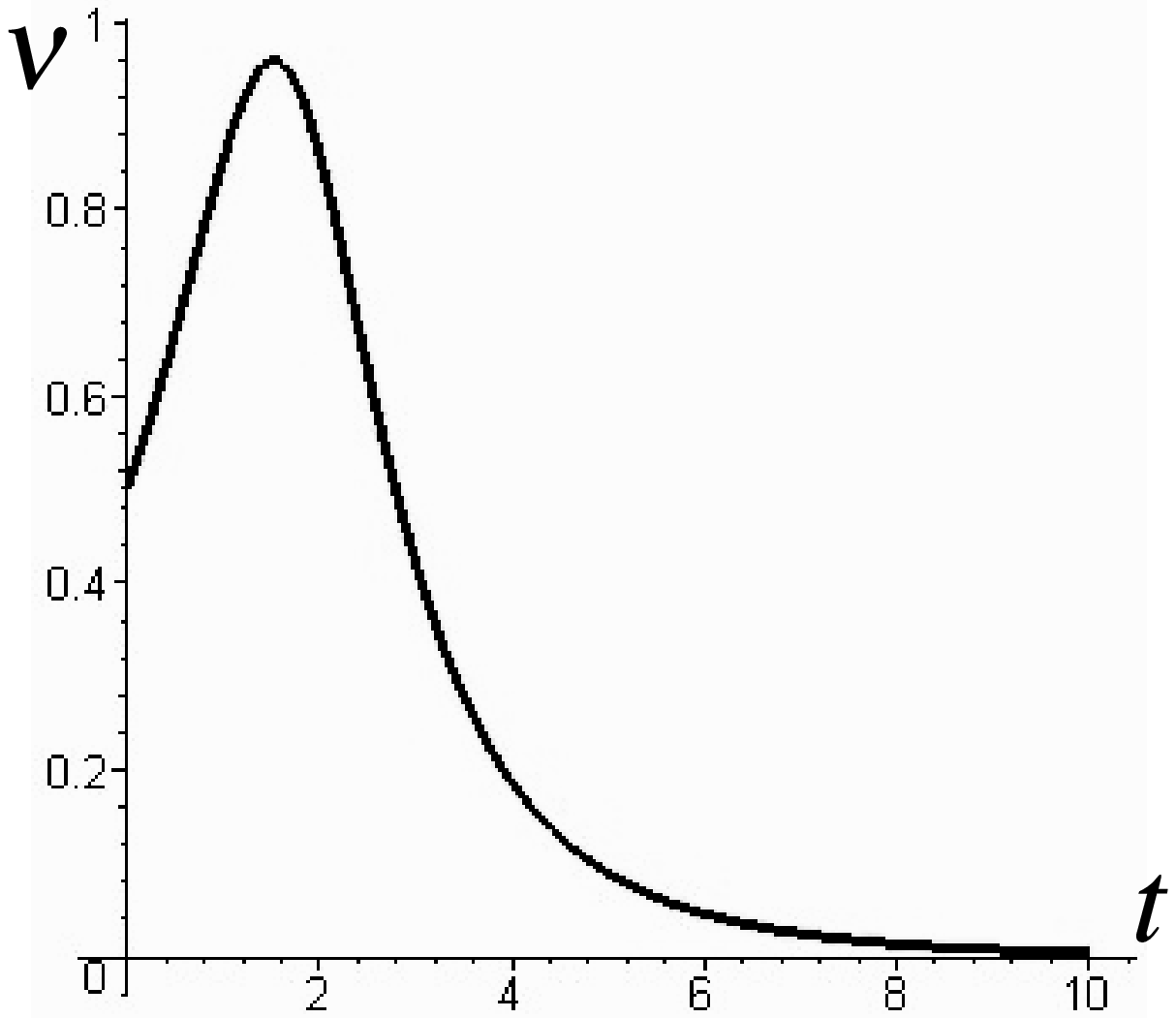}{0.25}{Evolution of function inverse to volume
scale}{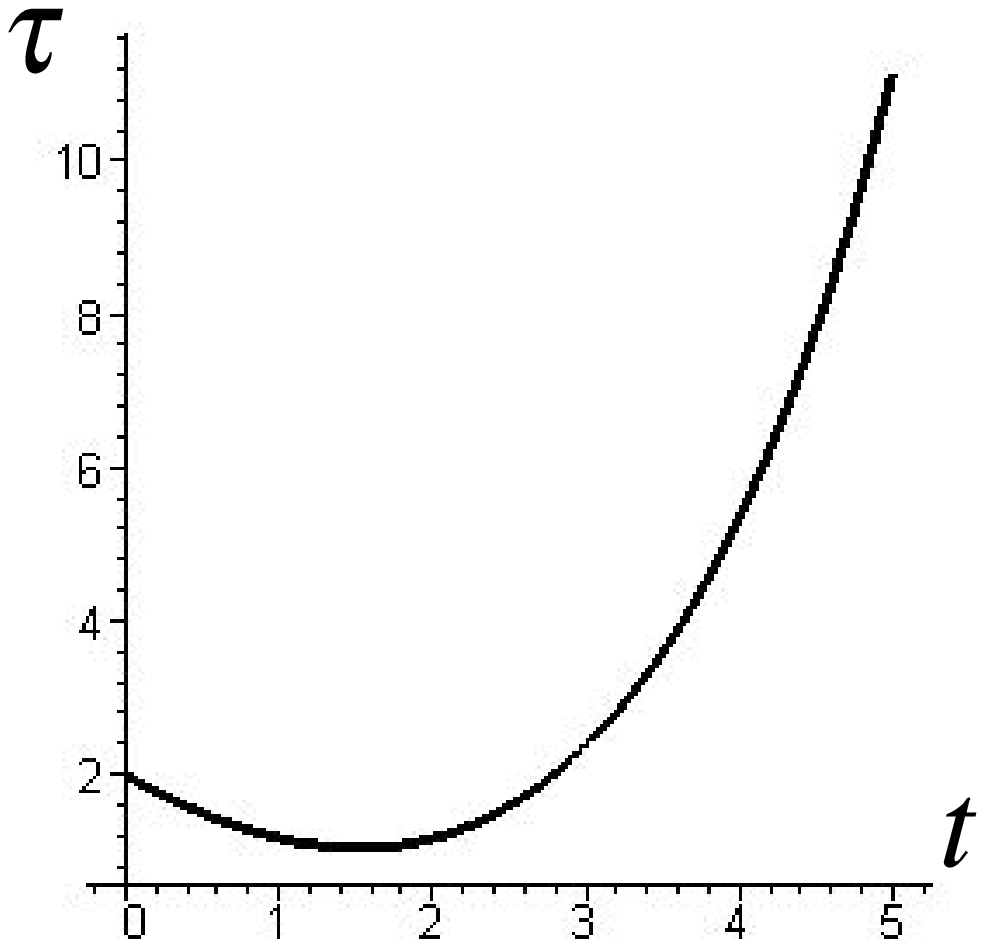}{0.20}{Evolution of volume scale}{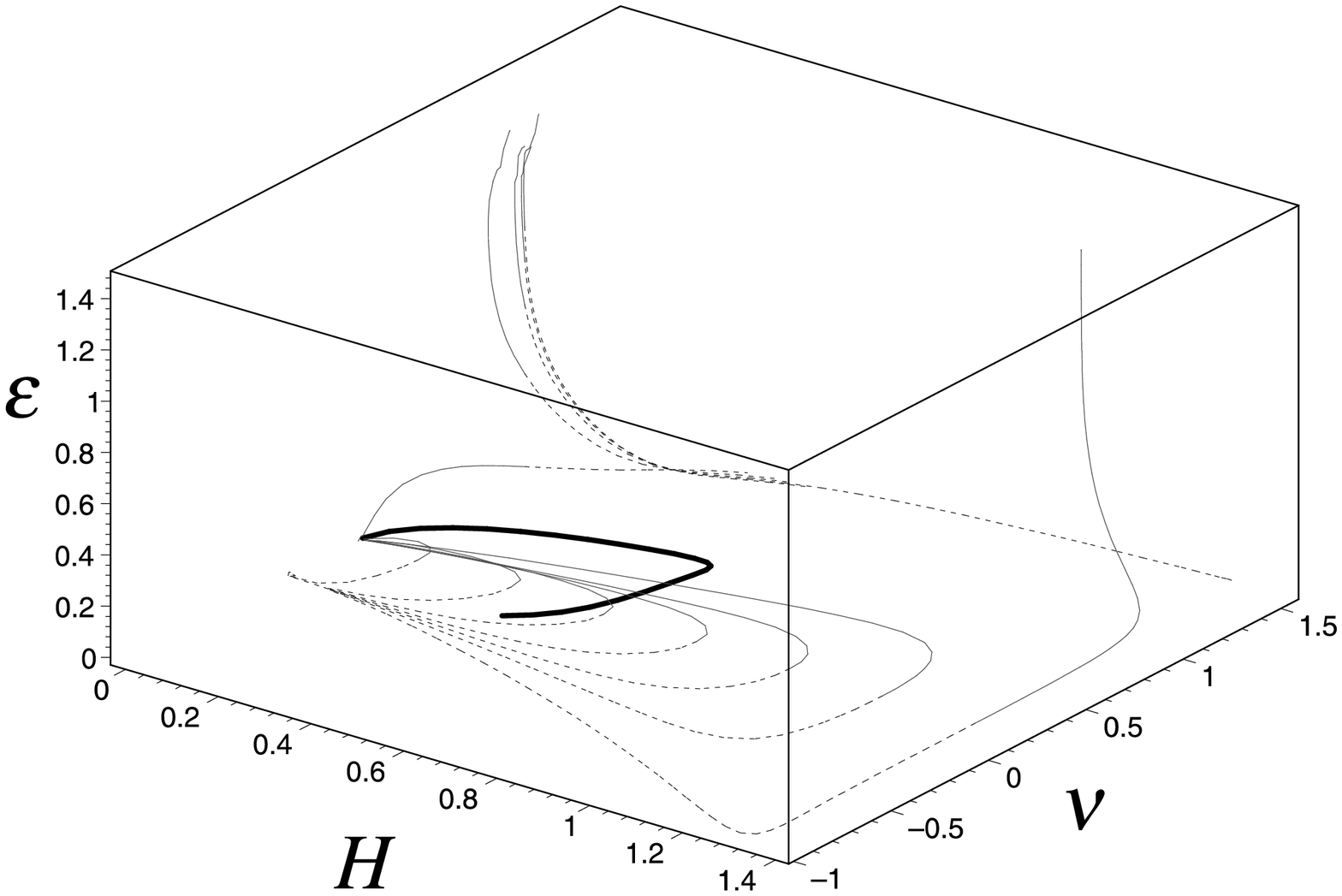}{0.20}{3D view
in $\nu,H,\ve$ space}


               \section{Conclusion}
Recently a self consistent system of nonlinear spinor and
gravitational fields in the framework of Bianchi type-I
cosmological model filled with viscous fluid was considered by one
of the authors \cite{saharrp,grqcnlsp}. The spinor filed
nonlinearity is taken to be some power law of the invariants of
bilinear spinor forms, namely $I = S^2 = (\bp \p)^2$ and $J = P^2
= (i \bp \gamma^5 \p)^2$. Solutions to the corresponding equations
are given in terms of the volume scale of the BI space-time, i.e.,
in terms of $\tau = a b c$, with $a,b,c$ being the metric
functions. This study generates a multi-parametric system of
ordinary differential equations \cite{saharrp,grqcnlsp}. Given the
richness of the system of equations in this paper a qualitative
analysis of the system in question has been thoroughly carried
out. A complete qualitative classification of the mode of
evolution of the universe given by the corresponding dynamic
system has been illustrated. In doing so we have considered all
possible values of the problem parameters independent to their
physical validity and graphically presented the most
distinguishable in our view results.


\newcommand{\hnl}{\htmladdnormallink}

\end{document}